\documentclass[prl,twocolumn,showpacs,preprintnumbers,amsmath,amssymb]{revtex4}

\usepackage{graphicx}% Include figure files
\usepackage{dcolumn}% Align table columns on decimal point
\usepackage{bm}% bold math

\begin{document}

\title{Feshbach Molecules in a One-dimensional Optical Lattice}
\author{N. Nygaard}
\author{R. Piil}
\author{K. M\o lmer}
\affiliation{Lundbeck Foundation Theoretical Center for Quantum System Research, Department of Physics and Astronomy, University of Aarhus, DK-8000 {\AA}rhus C, Denmark}

\date{\today}% It is always \today, today,
             %  but any date may be explicitly specified

\begin{abstract}
We present the theory of a pair of atoms in a one-dimensional optical lattice interacting via a narrow Feshbach resonance.  Using a two-channel description of the resonance, we derive analytic results for the scattering states inside the continuum band and the discrete bound states outside the band. We identify a Fano resonance profile, and the survival probability of a molecule when swept through the Bloch band of scattering states by varying an applied magnetic field. We discuss how these results may be used to investigate the importance of the structured nature of the continuum in experiments.    
\end{abstract}

% PACS, the Physics and Astronomy
                             % Classification Scheme.
%\keywords{Suggested keywords}%Use showkeys class option if keyword
                              %display desired
\maketitle

Ultracold atoms offer a compelling system for the study of fundamental physics, due to a large array of experimental tools allowing precision control of the state of the system. Two such tools are optical lattices~\cite{Bloch2007} and Feshbach resonances~\cite{Kohler2006}. With an optical lattice the atoms are trapped in a periodic light potential, while a Feshbach resonance is used to tune the interactions between the atoms. Recently, these tools have made interesting two-body~\cite{Moritz2005,Winkler2006,Bergeman2003,Orso2005,Grupp2007,Piil2007} and many-body~\cite{Carr2005,Zhou2005,Duan2005} states attainable.  

In this paper, we present an analytical description of the interplay between molecular binding and quantum motion in periodic potentials.
We consider a pair of atoms in a one-dimensional optical lattice near a narrow Feshbach resonance. Such  one-dimensional geometries are realized  
in a three-dimensional optical lattice when the depth of the transverse cosine potential is much larger than the longitudinal lattice depth~\cite{Moritz2005,Winkler2006}. 
In a discrete lattice model~\cite{Winkler2006,Piil2007}, we include on-site interactions, and consider the physics in the lowest Bloch band with nearest neighbor tunneling in the lattice. A two-channel model, in which  a bound state in a closed channel is coupled to the continuum, describes our Feshbach resonance~\cite{Kohler2006,Nygaard2006}. The formal description of our system bears a strong resemblance to Fano-Anderson theories of autoionizing states in atoms~\cite{Fano1961} and  localized electron states in solids~\cite{Anderson1961}.

The distinguishing feature of the lattice system is the structured nature of the continuum, which is divided into Bloch bands. As we are restricting our model to the lowest energy band, it means that the continuum has an upper edge. This has interesting and observable consequences on the interplay between bound and scattering states. Two of these, the transmission Fano profile and the survival probability of a molecule when swept through the continuum by varying an applied magnetic field across the resonance, are explicitly derived, and we discuss how they may be probed in current experiments.  

Introducing center of mass, $Z=(z_1+z_2)/2$, and relative, $z=z_1-z_2$, coordinates the lattice allows a separable ansatz for the relative and center of mass motion:
\begin{equation}
\psi(z_1,z_2)=e^{iKZ} \psi_K(z).
\label{sep_ansatz}
\end{equation}
The first Brillouin zone for the center of mass momentum, $K$, runs from $-\pi/a$ to $\pi/a$, where $a$ is the lattice spacing. 
The Hamiltonian for the relative motion in the open channel, $H^{\rm{op}}=[-2J\Delta_z^K+E_K+U\delta_{z,0}]\equiv H_0+U\delta_{z,0}$, contains the discrete Laplacian 
\begin{equation}
\Delta_z^K \psi_K(z) = -\frac{E_K}{4J} [\psi_K(z+a)+\psi_K(z-a)-2\psi_K(z)],
\label{lattice_laplacian}
\end{equation}
and the center of mass energy, $E_K = -4J \cos\left(Ka/2\right)$, in addition to the interaction between the atoms in the open channel. The on-site interaction in the entrance channel is $U$.
We take the zero of energy to be the middle of the band, which spans the energy interval $[-|E_K|,|E_K|]$. 

%The eigenstates of $H_0$ are Bloch waves, $\langle z|k\rangle=\sqrt{a/2\pi}\exp({ikz})$, with energies $\epsilon_K(k)=E_K\cos(ka)$. We note that $\epsilon_K(k)$ is the sum of single-particle energies, $-2J[\cos(Ka/2+ka)+\cos(Ka/2-ka)]$. The Green's function for free atoms on the lattice may be found from
%\begin{equation}
%G^0_K(E,z) = \int_{-\pi/a}^{\pi/a}  \frac{dk}{2\pi} \frac{ae^{ikz}}{E-E_K\cos(ka)+i\eta},
%\label{G0_int}
%\end{equation}
%where the infinitesimal $\eta>0$ ensures the correct boundary conditions. 
%The integral is over relative quasi-momenta. 
%Due to the delta function interactions we only need to consider $G_K^0(E,z)\equiv \langle z|\hat{G}_K^0(E)|0\rangle$.

The interacting Green's function is found by explicitly solving the Dyson equation, $\hat{G}^U_K(E)=\hat{G}_K^0(E)+\hat{G}_K^0(E)\hat{U}\hat{G}^U_K(E)$. In coordinate space the result has the simple form
\begin{equation}
G^U_K(E,z) \equiv \langle z|\hat{G}^U_K(E)|0\rangle = \frac{G^0_K(E,z)}{1-UG^0_K(E,0)},
\label{G_z}
\end{equation}
where $G^0_K(E,z)\equiv \langle z|\hat{G}_K^0(E)|0\rangle$ is the Green's function for the relative motion of two non-interacting atoms on the lattice, which for  energies inside the band $(|E|<|E_K|)$ is propagating
\begin{equation}
G^0_K(E,z) = -\frac{i\exp(ip|z|)}{\sqrt{E^2_K-E^2}},
\label{G0_z_in}
\end{equation}
with $pa=\cos^{-1}(E/E_K)$, while it falls of exponentially outside the band ($|E|>|E_K|$):
\begin{equation}
G^0_K(E,z) = {\rm{sign}}(E)\frac{\exp(-\kappa|z|)}{\sqrt{E^2-E^2_K}}[-{\rm{sign}}(E)]^{z/a}.
\label{G0_z_out}
\end{equation}
Here $\kappa a=\cosh^{-1}|E/E_K|$. For energies above the band the sign of the $G^0_K(E,z)$ alternates between lattice sites.
The pole of $G^U_K(E,z)$ at $1=UG_K^0(E_b^0,0)$ gives the energy of the bound pair state; $E_b^0={\rm{sign}}(U)\sqrt{E_K^2+U^2}$.
For $U>0$ the pole constitutes the repulsively bound pair studied recently~\cite{Winkler2006}.

In the following we derive the resonant scattering properties of the atoms and characterize the decay of molecular states in a two-channel model. We suggest avenues for the experimental investigation of these effects. 

Our aim is to solve the discrete coupled channel equations for the relative motion on  the lattice:
\begin{subequations}
\begin{equation}
H^{\rm{op}}\psi^{\rm{op}}_K(z) + W\delta_{z,0}\psi_K^{\rm{cl}}(z) 
= E \psi^{\rm{op}}_K(z), 
\label{coupled_eq1}
\end{equation}
\begin{equation}
H^{\rm{cl}}\psi^{\rm{cl}}_K(z) + W\delta_{z,0}\psi_K^{\rm{op}}(z) 
= E \psi^{\rm{cl}}_K(z),
\label{coupled_eq2}
\end{equation}
\end{subequations}
with a delta function coupling between the channels,
%\begin{equation}
$\langle z, \, {\rm{op}} |\hat{W}|z' , \, {\rm{cl}} \rangle = W \delta_{z,0}\delta_{z,z'}$.
%\label{W_diag}
%\end{equation}

From the spectrum of $H^{\rm{cl}}$ we only retain the closed channel bound state, $|\phi_{\rm{res}}\rangle$, which causes the resonance. It is detuned by an energy $E_{\rm{res}}(B,K)$ from the open channel band center and unit normalized, $\langle \phi_{\rm{res}}|\phi_{\rm{res}}\rangle=1$. Hence we treat the closed channel Green's function, $\hat{G}_K^{\rm{cl}}(E)=[E-\hat{H}^{\rm{cl}}]^{-1}$, within a single-pole approximation
\begin{equation}
\hat{G}_K^{\rm{cl}}(E) \approx  \frac{|\phi_{\rm{res}}\rangle \langle \phi_{\rm{res}}|}{E-E_{\rm{res}}(B,K)}.
\label{G_cl_approx}
\end{equation}
The bare resonance energy varies linearly with the applied magnetic field, the slope being the difference between the closed and open channel magnetic moments, $\Delta\mu$. To simplify the notation we introduce an effective coupling $\langle z, \, {\rm{op}} |\hat{W}|\phi_{\rm{res}}\rangle = {\mathcal{W}}\delta_{z,0}$. Both ${\mathcal{W}}$ and $U$ may be obtained from free space scattering parameters by integrating over 3D tight-binding orbitals, and can be tuned by adjusting the transverse lattice.

A numerical analysis based on perturbation theory shows that our restriction to the lowest band is valid provided the dimensionless coupling strength $({\rm{amu}}/m)(a_{\rm{bg}}/100a_0)(\Delta\mu/\mu_{\rm{B}})(\Delta B/{\rm{G}})$ is small compared with  $0.01(V_0^{\|}+3)(V_0^{\perp})^{-0.6}$, where $m$ is the atomic mass, $a_{\rm{bg}}$ is the background scattering length, and $\Delta B$ is the magnetic field width of the resonance. The depths of the cosine optical lattice in the transverse, $V_0^{\perp}$, and longitudinal, $V_0^{\|}$, directions are both measured in recoil units. The scaling factors are the atomic mass unit, amu, the Bohr radius, $a_0$, the Bohr magneton, $\mu_{\rm{B}}$, and Gauss, $G$. If this criterion is not met, coupling to higher bands may not be neglected~\cite{Stoof_papers,Diener2006}.

The scattering amplitude, $f(E,K)$, is deduced from the open channel component of the asymptotic solutions to Eqs. (\ref{coupled_eq1},\ref{coupled_eq2}). It can be written as $f(E,K)=f_{\rm{bg}}(E,K)+f_{\rm{res}}(E,K)$, where the background contribution, $f_{\rm{bg}}(E,K)=UG_K^0(E,0)/[1-UG_K^0(E,0)]$ is associated with scattering in the absence of the coupling and has been described in \cite{Winkler2006}, while the resonant part is 
\begin{equation}
f_{\rm{res}}(E,K)=\frac{{\mathcal{W}}^2G_K^0(E,0)/[1-UG_K^0(E,0)]^2}{E-E_{\rm{res}}(B,K)-G^U_K(E,0){\mathcal{W}}^2}.
\label{f_res}
\end{equation}
The scattering states cover the same range of energies, $\epsilon_K(k)=E_K\cos(ka)$, as the non-interacting pairs, and  
outside the band the poles of (\ref{f_res}), $E_b$, determined by 
\begin{equation}
\left[\sqrt{E_b^2-E_K^2}-U{\rm{sign}}(E_b)\right]=\frac{{\mathcal{W}}^2{\rm{sign}}(E_b)}{E_b-E_{\rm{res}}(B,K)},
\label{E_bind}
\end{equation}
correspond to discrete bound states of the coupled system. The bound state energies may be identified as a subset of the solutions to a quartic equation. The apparent pole at the uncoupled bound state energy, $E_b^0$, has no weight, since the residues from the background and resonant parts of $f(E,K)$ cancel. 

The size of the bare resonance state is much smaller than the lattice spacing, and thus confined to a single lattice site, the bare molecules effectively experience a lattice twice as deep as the one in which the atoms move. Hence the molecular tunneling rate is greatly suppressed with respect to its atomic counterpart, and we neglect in the following the modulation of the resonance energy by the center of mass motion, $E_{\rm{res}}(B,K)\approx E_{\rm{res}}(B)$.

\begin{figure}[!h] %  figure placement: here, top, bottom, or page
   \centering
   \includegraphics[width=8.6cm]{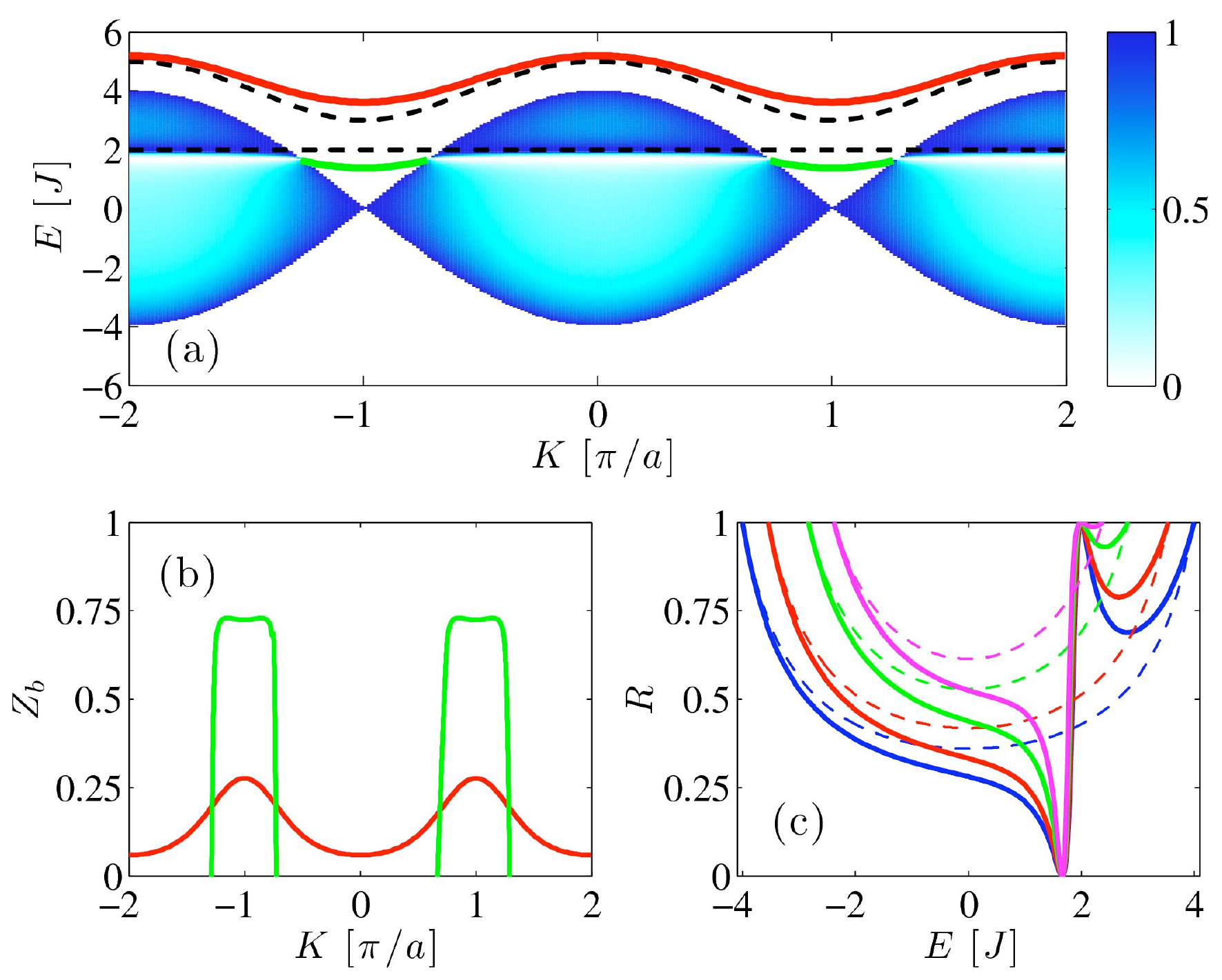} 
   \caption{(Color online) Scattering and bound states in an extended zone scheme for $E_{\rm{res}}=2J$ (a). The colormap inside the band illustrates the reflection coefficient, $R$, ranging from 0 (white) to 1 (blue). The solid lines are the bound state energies of the coupled system, while the dashed lines indicate the diabatic, uncoupled bound states, with the horizontal line giving $E_{\rm{res}}$, and the other corresponding to $E_b^0$. The on-site interaction and the coupling strength are $U=3J$ and ${\mathcal{W}}=1J$, respectively. (b): closed channel fraction of the bound states, Eq. (\ref{Z}). (c): resonance profiles (solid lines) for $E_{\rm{res}}=2J$ at $Ka/\pi=0$, 0.3, 0.5, and 0.6 (from below). The dashed lines are the corresponding curves for the uncoupled case, where there is no resonance.}
   \label{fig1}
\end{figure}

\begin{figure}[!h] %  figure placement: here, top, bottom, or page
   \centering
   \includegraphics[width=8.6cm]{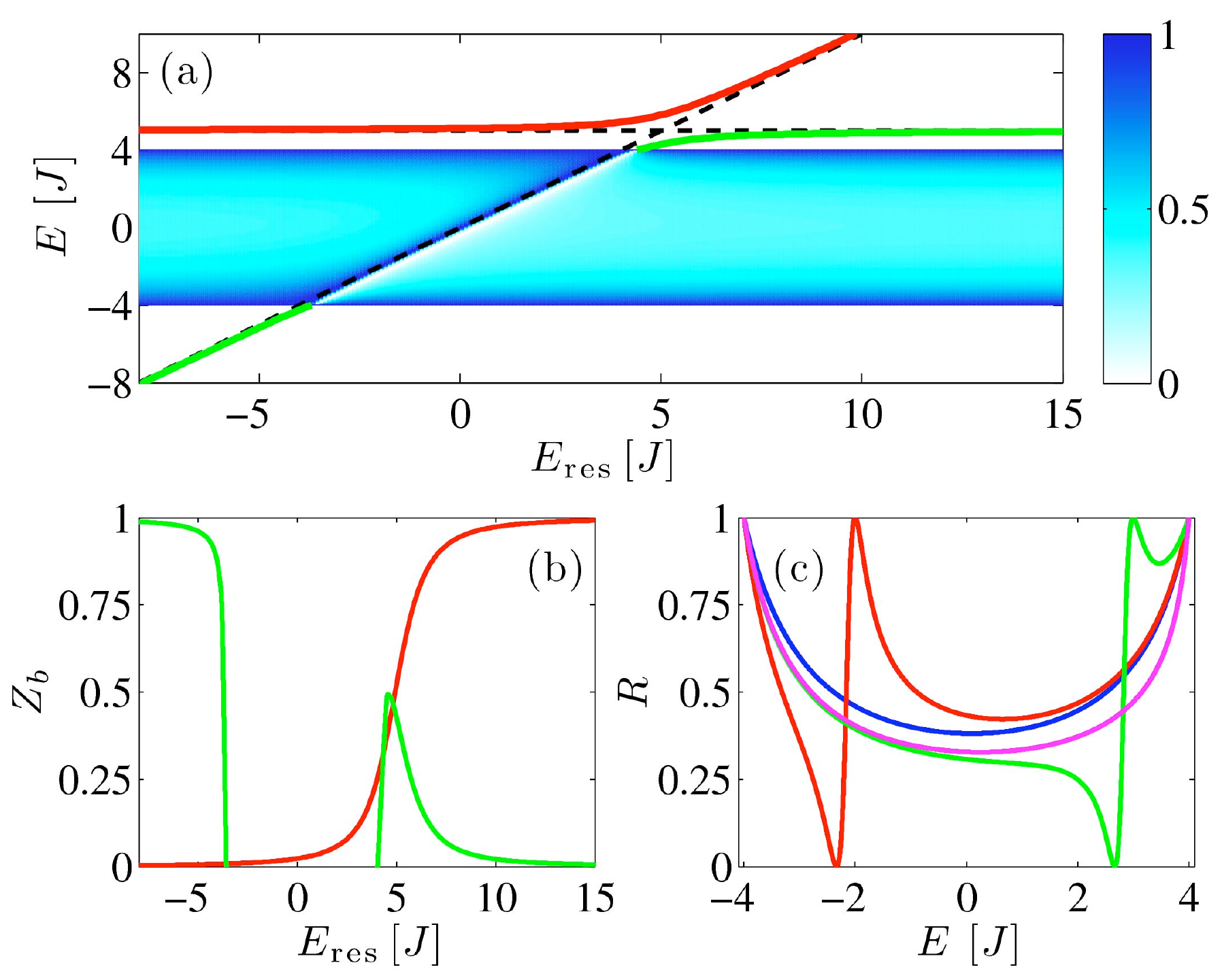} 
   \caption{(Color online) (a):  as Fig.~\ref{fig1}(a), but plotted as a function of $E_{\rm{res}}$ for fixed $Ka/\pi=0$. The horizontal dashed line indicate the energy of the bound pair for ${\mathcal{W}}=0$, and the slanted dashed line gives the bare resonance energy, $E_{\rm{res}}$. (b): closed channel fraction of the bound states, Eq. (\ref{Z}). (c): resonance profiles for $Ka/\pi=0$ at $E_{\rm{res}}/J=-2$, -7, 5, 3 (from the top at $E/J=0$).}
   \label{fig2}
\end{figure}

Figures~\ref{fig1}(a) and~\ref{fig2}(a)  show the variation of the scattering properties and the energies of the discrete bound states with the center of mass momentum and the bare resonance energy, respectively. The scattering is represented through the reflection coefficient, $R=|f|^2$, which ranges from 0 (no scattering) to 1 (unitarity). The unitary scattering at the band edges is evidence of the van Hove singularities in the density of states, ${\rm{Im}}G_K^0(E,0)$.
In the uncoupled case, where ${\mathcal{W}}=0$, two diabatic bound states exist, as indicated by the dashed lines in Figures~\ref{fig1}(a) and~\ref{fig2}(a). The bare resonance state (horizontal line in Fig.~\ref{fig1}(a), sloped line in Fig.~\ref{fig2}(a)) gives rise to the clearly visible scattering resonance  where it cuts through the band. The bound state pole, $E_b^0$, of the open channel Green's function (\ref{G_z}) is represented by the sinusoidal and horizontal dashed lines in Figures~\ref{fig1}(a) and \ref{fig2}(a), respectively.
For a finite coupling the two bound states repel each other, leading to the avoided crossing in Fig.~\ref{fig2}. 

The coupling of the bare resonance state to the continuum gives it a finite width, whenever it is embedded in the band. However, the resonance only exists for a range of center of mass motion Bloch states for a given magnetic field, as shown in Fig.~\ref{fig1}(a)~\cite{Grupp2007}. Outside this range (and between higher bands~\cite{Syassen2007}) 
%the Feshbach molecule (lower solid line in Fig.~\ref{fig1}(a)) is 
there is instead
a true bound state of the system.

The analytic shape of the resonance is described by a Fano profile in the transmission probability
\begin{equation}
T=1-R=T_{\rm{bg}}\frac{(\epsilon+q)^2}{\epsilon^2+1},
\label{Fano}
\end{equation}
arising due to interference between open and closed channel scattering. The reduced variables $\epsilon=2(E-E_{\rm{res}}-\Delta)/(\hbar\Gamma)$ and $q=2\Delta/\hbar\Gamma$ contain shift and width functions related to the the real and imaginary parts of the molecular self energy: $\Delta(E,K)-i\hbar\Gamma(E,K)/2=\langle \phi_{\rm{res}}|\hat{W}\hat{G}^U_K(E)\hat{W}|\phi_{\rm{res}}\rangle = -{\mathcal{W}}^2(U+i\sqrt{E_K^2-E^2})/(E_K^2+U^2-E^2)$, while the background transmission coefficient is $T_{\rm{bg}}=1/(1+q^2)$. Note, that $T$ assumes a simple Breit-Wigner profile, when the background interaction is ignored ($U=0$, $T_{\rm{bg}}=1$)~\cite{Grupp2007}.

We are aware that the lowest band approximation under which our calculations are valid is
rather restrictive. However, to lowest order in the atom-molecule coupling the inclusion of higher bands 
only gives rise to real corrections to the molecular self-energy, modifying the shift, $\Delta$, of the resonance energy,
while leaving the lifetime unchanged. Furthermore, the shift will be only weakly energy dependent, because the considered energies are well
separated from the higher bands. The Fano resonance profile in the transmission is thus expected to be essentially unchanged if higher bands are included.

The Fano profile (\ref{Fano}) describes the scattering states of the system, but it
also parametrizes the probability of transitions from a different initial
state to the energy eigenstates inside the band. In the same way as Fano's theory was
originally used to describe photoionization via autoionizing
states \cite{Fano1961}, Eq. (\ref{Fano}) and Figures~\ref{fig1}(c) and~\ref{fig2}(c) describe the
probability of RF- or photodissociative transitions from deeper bound states
in the molecule. 
%In addition to its spectroscopic investigation by way of the atomic dissociation signal, the Fano
%dip in this probability may have interesting applications for the control of coupled atomic and molecular matter waves~\cite{Harris2002}.

We now turn to the bound states of the coupled system. Their wavefunctions have open and closed channel parts given by $\psi_b^{\rm{op}}(z)=\sqrt{Z_b}{\mathcal{W}}G^U_K(E_b,z)$ and $\psi_b^{\rm{cl}}(z)=\sqrt{Z_b}\delta_{z,0}$, respectively. 
From normalization it follows that the closed channel fraction of the bound state at energy $E_b$ is
\begin{equation}
Z_b = \left[ 1+\frac{{\mathcal{W}}^2}{[1-UG^0_K(E_b,0)]^2}\frac{|E_b|}{(E_b^2-E_K^2)^{3/2}}\right]^{-1},
\label{Z}
\end{equation}
which is plotted in Figures~\ref{fig1}(b) and~\ref{fig2}(b). For the Feshbach molecule $Z_b$ vanishes exactly when the state enters the continuum, becoming the scattering resonance.  

In momentum space the component of the coupled channels bound state in the open channel is 
\begin{equation}
\phi_b^{\rm{op}}(k) = \sqrt{\frac{aZ_b}{2\pi}}{\mathcal{W}}\frac{{\mathcal{G}}_K^0(E_b,k)}{1-UG_K^0(E_b,z=0)},
\label{phi_mom}
\end{equation}
while $\phi_b^{\rm{cl}}(k)=\sqrt{aZ_b/2\pi}$ for all $k$.
The non-interacting quasi-momentum Green's function is ${\mathcal{G}}_K^0(E,k)=[E-E_K\cos(ka)+i\eta]^{-1}$. For $|K|<\pi/a$ and for $E_b>0$ ($E_b<0$) the bound state function $\phi_K^{\rm{op}}(k)$ is peaked at the edges (center) of the Brillouin zone. This reflects that $\phi_K^{\rm{op}}(k)$ is predominantly comprised of the $k$-states with energies in proximity of $E_b$. For $E_b>0$ ($E_b<0$) this corresponds to the states near the top (bottom) of the band~\cite{Piil2007}. 
Thus the location of the peaks makes it possible to distinguish between bound states lying above or below the continuum in  time of flight~\cite{Winkler2006}. 
The height of the peaks are reduced as $K$ approaches $\pm\pi/a$, where all quasi-momentum states are degenerate in our model.  

%\begin{figure}[!h] %  figure placement: here, top, bottom, or page
%   \centering
%   \includegraphics[width=\columnwidth]{Fig3.eps} 
%   \caption{Open channel components of bound state wavefunctions for $E_{\rm{res}}=-4J$, where the Feshbach molecule (blue) is situated just below the band. The repulsively bound pair state (red), is peaked at the edge of the Brillouin zone, and its sign alternates from one site to the next. The first column shows the momentum space wavefunction (\ref{phi_mom}) in units of $\sqrt{a/2\pi}$, the second column $\psi^{\rm{op}}_b(z)$. Same parameters as in Fig.~\ref{fig1}.}
%   \label{fig3}
%\end{figure}

%The bound state wavefunctions in both momentum and coordinate space are illustrated in Fig.~\ref{fig3} for $E_{\rm{res}}=-4J$, where Feshbach molecules at rest are just barely bound. The size of the bound state
%\begin{equation}
%\langle |z| \rangle=\frac{a}{2}\frac{Z_b{\mathcal{W}}^2}{[1-UG_K^0(E_b,0)]^2}\frac{E_K^2}{(E^2_b-E^2_K)^2}, 
%\end{equation}
%diverges when $E_b\rightarrow \pm|E_K|$.

\begin{figure}[!h] %  figure placement: here, top, bottom, or page
   \centering
   \includegraphics[width=\columnwidth]{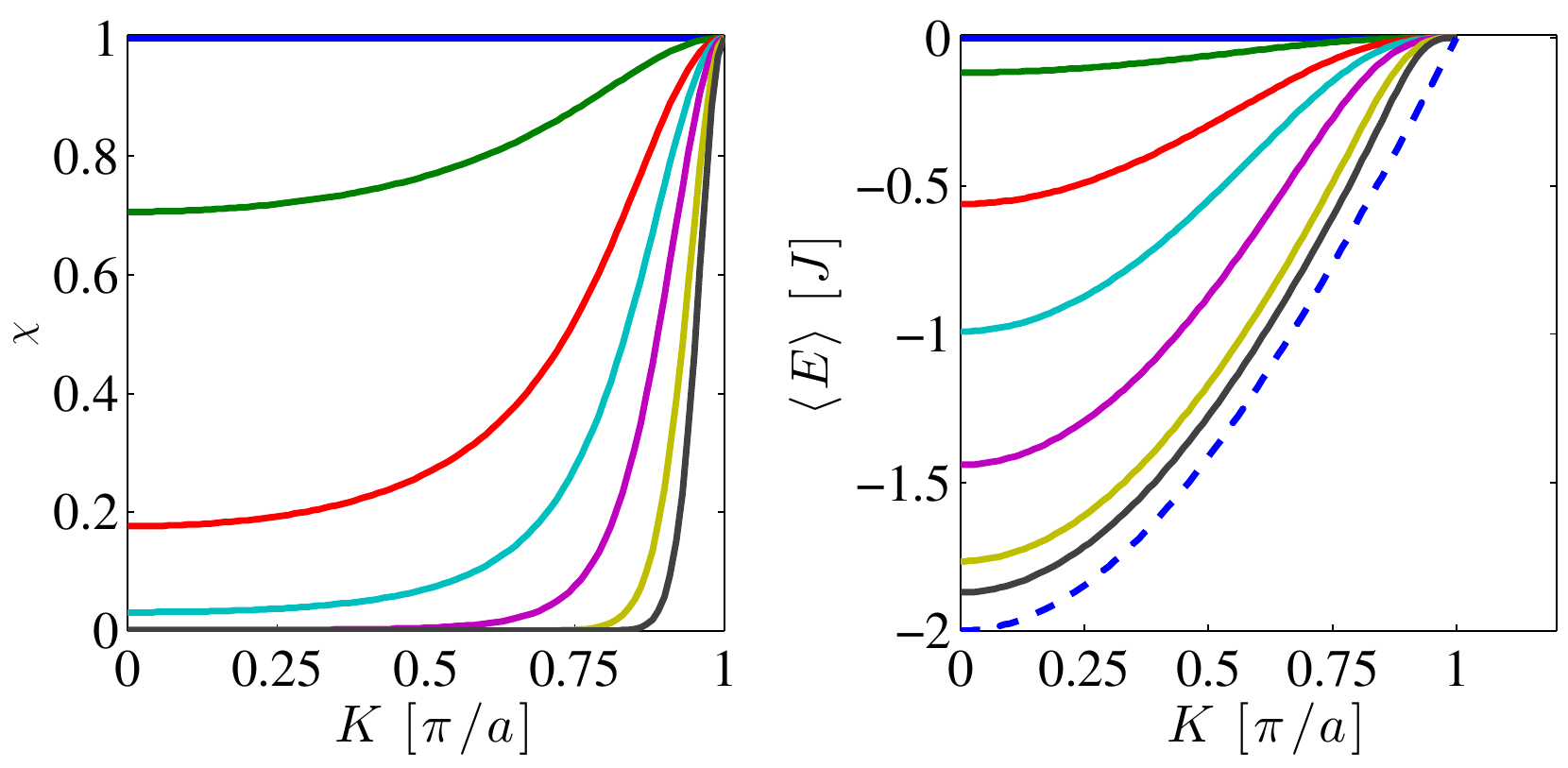} 
   \caption{Remaining molecule fraction after a linear sweep through the band (left) and average dissociation energy per atom (right) for $\alpha$=0, 0.1, 0.5, 1, 2, 5, 10 (from the top). $U=2J$. For fast ramps all molecules survive, but as the ramp speed is decreased more molecules decay and the dissociation energy approaches half the lower band edge (dashed line).}
   \label{fig3}
\end{figure}

%In summary, we have identified the energy eigenstates of the coupled channels system and
%a number of interesting features due to the interplay of the Feshbach
%resonance and the motion of the particles in a periodic potential. 
%Let us
%now conclude with a discussion of the most striking properties of the system and
%proposals for experimental investigation. 
At a given energy inside the band the Feshbach molecules are unstable. Their dissociative decay width is $\hbar\Gamma(E,K)$ given above.
For a sweep of the magnetic field across resonance, starting with the resonance energy tuned below the band, the molecular population follows a simple rate equation, $dm/dt=-\Gamma(E(t))m(t)$, provided the sweep is fast enough that recombination may be neglected. 
For a linear sweep, where $dE/dt=\Delta\mu\dot{B}$, the remaining molecule fraction can be found analytically as a function of the  energy inside the band: $m(E)=\exp\lbrace
-2\alpha
[
\left( \sin^{-1} E/|E_K|+\pi/2 \right) -C ( \tan^{-1}CE/|E_K|\sqrt{1-(E/|E_K|)^2}+\pi/2 ) 
]
\rbrace$,
where $\alpha={\mathcal{W}}^2/\Delta\mu\dot{B}\hbar$, $\beta=U/E_K$, and $C=\sqrt{\beta^2/(1+\beta^2)}$. A unique feature of the lattice system, is that the continuum has an upper edge, and for a sweep ending above the band the final molecule fraction has the simple form
\begin{equation}
\chi\equiv m(|E_K|)=\exp\left[-2\pi \alpha \left( 1 - \sqrt{\frac{\beta^2}{1+\beta^2}} \right) \right],
\label{mE_end}
\end{equation}
depending only on the two constants $\alpha$ and $\beta$. This expression is plotted in the left hand panel of Fig.~\ref{fig3} for $U=2J$. Again, this result is expected to be modified only slightly by the inclusion of higher bands.

In the case of vanishing background interaction, $U=0$, the survival probability assumes a Landau-Zener form, in which $\alpha$ plays the role of the adiabaticity parameter. We also note that in the limit $U\rightarrow \pm\infty$ all molecules survive the passage through the band. 
The limit $\alpha\rightarrow 0$, corresponding to sweeps instantaneous on the time scale set by the coupling, also leaves the molecules intact. Finally, $\chi\rightarrow 1$ at $K=\pi/a$ due to the vanishing width of the band at the center of mass zone boundary. This is an artifact of omitting tunneling beyond nearest neighbor. As more distant tunneling amplitudes are incorporated the band widens at the zone boundary and an extra van Hove singularity develops in the density of states~\cite{Piil2007}. Lastly, we note that $\chi$ is {\textit{independent}} of the sign of $U$.

The average energy of the dissociated atoms 
is plotted in the right hand panel in Fig.~\ref{fig3}. As a smaller fraction of the molecules survive the passage through the band the dissociation energy per atom approaches half the lower band edge, $-|E_K|/2=-2J\cos(Ka/2)$. 
Molecule dissociation energies have been measured in free space~\cite{Mukaiyama2004,Durr2004}, and the energy distribution after the sweep through the energy band may be identified by studying the ballistic expansion of the gas after the atoms are released from the lattice on a time scale, admitting adiabatic mapping of the quasi-momentum states to plane wave momentum states~\cite{Bloch2007}. 

N. N. acknowledges financial support by the 
Danish Natural Science Research Council.

\end{document}